\documentclass[twocolumn,showpacs,prl]{revtex4}
\usepackage{bm}
\usepackage{epsfig}
\newcommand{\nix}[1]{}

\begin{document}

\title{
Observation of Spin Relaxation Anisotropy in Semiconductor Quantum
Wells}
\author{N.S.~Averkiev, L.E.~Golub, A.S.~Gurevich, V.P.~Evtikhiev, V.P.~Kochereshko, A.V.~Platonov, A.S.~Shkolnik, and
Yu.P.~Efimov}\altaffiliation[Address: ]{Institute of Physics, St.Petersburg State University,
Ulyanovskaya 1, Petrodvorets, 198504 St.~Petersburg, Russia}
\affiliation{A.F.~Ioffe Physico-Technical Institute, Russian
Academy of Sciences, 194021 St.~Petersburg, Russia}
%


\begin{abstract}
Spin relaxation of two-dimensional electrons in asymmetrical (001)
AlGaAs quantum wells are measured by means of Hanle effect. Three
different spin relaxation times for spins oriented along [110],
$[1\bar{1}0]$ and [001] crystallographic directions are extracted
demonstrating anisotropy of D'yakonov-Perel' spin relaxation
mechanism. The relative strengths of Rashba and Dresselhaus terms
describing the spin-orbit coupling in semiconductor quantum well
structures. It is shown that the Rashba spin-orbit splitting is
about four times stronger than the Dresselhaus splitting in the
studied structure.
\end{abstract}

\maketitle

Spintronics is at present time one of the most important areas of
the semiconductor physics for both fundamental research and
possible device applications~\cite{spintronicbook02}. The main
problem of spintronics is creation, registration and lifetime
control of carrier spin, especially in low-dimensional systems.
Therefore investigation of spin relaxation processes is now an
actual problem of the physics of semiconductor heterostructures.

The main mechanism of spin relaxation in GaAs based quantum wells
(QWs) is the D'yakonov-Perel' kinetic
mechanism~\cite{Dyakonov86p110}. It is caused by lack of inversion
centrum i)~in the bulk semiconductor of which the system is made
(bulk inversion asymmetry, or BIA), ii)~in the heterostructure
(structure inversion asymmetry, or SIA) and iii)~in the chemical
bonds at heterointerfaces (interface inversion asymmetry, or
IIA)~\cite{Dyakonov86p110,Bychkov84p78,Roessler02p313}. SIA can be
caused by an external electric field or by deformation, BIA and
IIA depend strongly on a size of carrier confinement. Therefore
spin relaxation times can be controlled by gate voltage or by
special heterostructure design.

In Ref.~\cite{Averkiev1999p15582}, anisotropy of spin relaxation
has been predicted for heterostructures grown along the axis
[001]. It has been theoretically shown that lifetimes of spin
oriented along the axes [110], $[1\bar{1}0]$ and [001] are
different. In particular, changing relation between SIA and BIA
one can achieve total suppression of relaxation for spin oriented
along one of $\langle 110 \rangle$ axes. (IIA in (001)-grown
structures is equivalent to BIA, therefore we will use a
generalized term `BIA' for both of them.) Detailed
calculations~\cite{Averkiev02pR271,FTP,Kainz03p075322} confirmed
that spin relaxation anisotropy exists in real semiconductor
heterostructures. Realization of such idea to control spin
relaxation times gives new opportunities for
spintronics~\cite{Schliemann03p146801}. However experimental
discovery of this effect is missed so far.

In this Letter, spin relaxation anisotropy in the plane of the QW
is observed. In order to demonstrate this effect, the structure
has been grown so that SIA and BIA are comparable in efficiency.
Note that systems where both SIA and BIA take place have been
studied in
Refs.~\cite{PikusPikus,Jusserand95p4707,Miller03p076807,BIASIA}
but spin relaxation times have not been investigated in such
structures.

The D'yakonov-Perel' spin relaxation mechanism consists in
electron spin precession around an effective magnetic field which
is caused by lack of inversion centrum in the system. The
corresponding Hamiltonian of spin-orbit interaction has the form
\begin{equation}\label{HSO}
    H = \hbar {\bm \sigma} \cdot {\bm \Omega}({\bm k}),
\end{equation}
where ${\bm \sigma}$ is a vector of Pauli matrices and ${\bm
\Omega}$ is a precession frequency dependent on the electron
quasimomentum $\bm k$.

The direction of the precession axis ${\bm \Omega}$ is determined
by the carrier momentum $\bm k$ and by kind of inversion
asymmetry. SIA generates the effective field oriented
perpendicular to $\bm k$. BIA results in the field which direction
depends on the angle between the momentum and crystallographic
axes. In (001) QWs both SIA and BIA produce effective magnetic
fields lying in the plane of the structure.  In the coordinate
system $x
\parallel [1\bar{1}0]$, $y \parallel [110]$ the precession frequences have the
form
\begin{equation}\label{Omega}
    {\bm \Omega}_{\rm SIA} = \alpha (k_y, -k_x), \:\:\:\:
    {\bm \Omega}_{\rm BIA} = \beta (k_y, k_x).
\end{equation}

If only one kind of inversion asymmetry is present, e.g. SIA, then
the precession frequency is the same for all momentum directions:
$|{\bm \Omega}| = \alpha k$, Eq.(\ref{Omega}). As a result, spin
relaxation times do not depend on spin orientation in the
structure plane~\cite{Dyakonov86p110}.

\begin{figure}[t]
\includegraphics[width=0.8\linewidth]{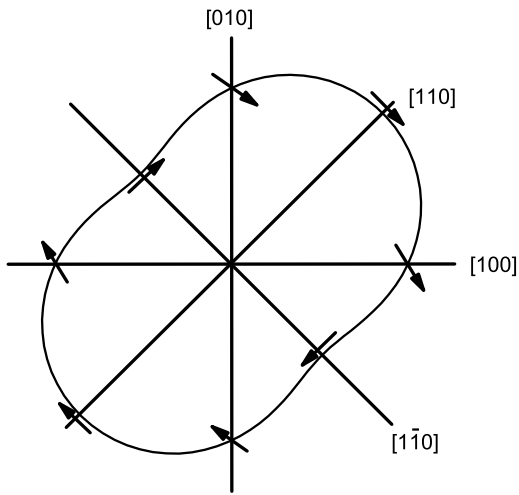}
\caption{Direction (arrows) and magnitude (solid line) of the
field ${\bm \Omega}({\bm k})$ in the ${\bm k}$-space at
$\alpha/\beta =4$.}
 \label{figOmega}
\end{figure}

If the system has both kinds of inversion asymmetry then the
effective magnetic field is a vector sum of the corresponding SIA
and BIA terms: ${\bm \Omega} = {\bm \Omega}_{\rm SIA} + {\bm
\Omega}_{\rm BIA}$. In this case the precession frequency depends
on the momentum direction of carriers~\cite{Silva92}:
$$|{\bm \Omega}| = k \sqrt{\alpha^2 + \beta^2 -
2\alpha\beta\cos{2\theta}},$$
where $\theta$ is an angle between $\bm k$ and the axis
$[1\bar{1}0]$. The angular dependence ${\bm \Omega}({\bm k})$ is
presented in Fig.~\ref{figOmega}. Due to anisotropy of ${\bm
\Omega}$ in the ${\bm k}$-space, the spin relaxation rate depends
on the spin orientation relative to crystallographic
axes~\cite{Averkiev1999p15582,Averkiev02pR271}. In particular, if
SIA and BIA strengths are identical ($|\alpha| = |\beta|$), then
the effective magnetic field is oriented along the same axis for
all directions of electron momentum. Therefore the spin oriented
along this direction (one of $\langle 110 \rangle$ axes) does not
relax at all. Two other spin components disappear with finite
rate. This means giant spin relaxation anisotropy.

Spin relaxation times are given by the following
expressions~\cite{Averkiev1999p15582}
\begin{equation}\label{tauz}
    {1 \over \tau_z} = C (\alpha^2 + \beta^2),
\end{equation}
for spin oriented along the growth axis, and
\begin{equation}\label{tauinplane}
    {1 \over \tau_\pm} = C {(\alpha \pm \beta)^2 \over 2}
\end{equation}
for spins oriented along $[1\bar{1}0]$ and [110] axes. Here $C$ is
a factor determined by temperature and momentum relaxation time
independent of spin-orbit interaction parameters.

In Eqs.~(\ref{Omega})-(\ref{tauinplane}) we neglect $\bm k$-cubic
terms in $\bm{\Omega}$. Since typical electron kinetic energy at
liquid nitrogen temperature relevant in the experiment is much
less than the energy of size quantization, the role of these terms
in spin relaxation is inessential~\cite{Averkiev02pR271,FTP}.

The spin relaxation times can be measured in time-resolved or
steady-state photoluminescence (PL) experiments. In the present
work, we used the method of PL depolarization by transverse
magnetic field (Hanle effect). The Hanle linewidth is determined
by a lifetime of spin oriented perpendicular to the magnetic field
$\bm B$. Therefore measuring PL circular polarization degree in
the geometries ${\bm B} \parallel [110]$ and ${\bm B} \parallel
[1\bar{1}0]$, one can extract the times $\tau_+$ and $\tau_-$.

The degree of circular polarization of radiation in these two
geometries has a Lorentzian form
$$
    P_{circ}({\bm B}) = {P_{circ}(0) \over 1 + (B/B_\pm)^2}.
$$
Here the halfwidths are given by
\begin{equation}\label{B_pm}
B_\pm = {\hbar \over g \mu_{\rm B} \sqrt{\tau_z \tau_\pm}},
\end{equation}
where $g$ is an electron factor Land\'e in the QW plane, and
$\mu_{\rm B}$ is the Bohr magneton. We suppose that spin
relaxation is much faster than radiation recombination:
$\tau_i \ll \tau_0$ ($i = z,+,-$; $\tau_0$ is the radiative
recombination time). In this model, under recombination of
electrons with heavy holes in the ground state, $P_{circ}(0) =
\tau_z/\tau_0$.

From~Eq.~(\ref{tauinplane}),~(\ref{B_pm}) follows the expression
for spin splittings
\begin{equation}\label{ratio} {B_+ - B_- \over B_+ + B_-} = \left|
{\beta \over \alpha} \right|.
\end{equation}
One can see that, even for dominance of one splitting over another
(e.g. $|\beta/\alpha| \ll 1$), spin relaxation anisotropy can be
registered experimentally because halfwidths of the Hanle curves
differ two times stronger than the spin splittings: $B_+ / B_-
\approx 1 + 2 \beta / \alpha$.

If one knows the $g$-factor and the Hanle contour halfwidths
$B_\pm$, one can determine the products $\tau_z \tau_+$ and
$\tau_z \tau_-$. For the D'yakonov-Perel' mechanism, the spin
relaxation rates obey the following relation
$$ {1 \over \tau_+} + {1 \over \tau_-} = {1 \over \tau_z}$$
- cf.~Eqs.~(\ref{tauz}),~(\ref{tauinplane}). This allows one to
determine all three spin relaxation times from the Hanle effect
measurements.

\begin{figure}[t]
\includegraphics[width=\linewidth]{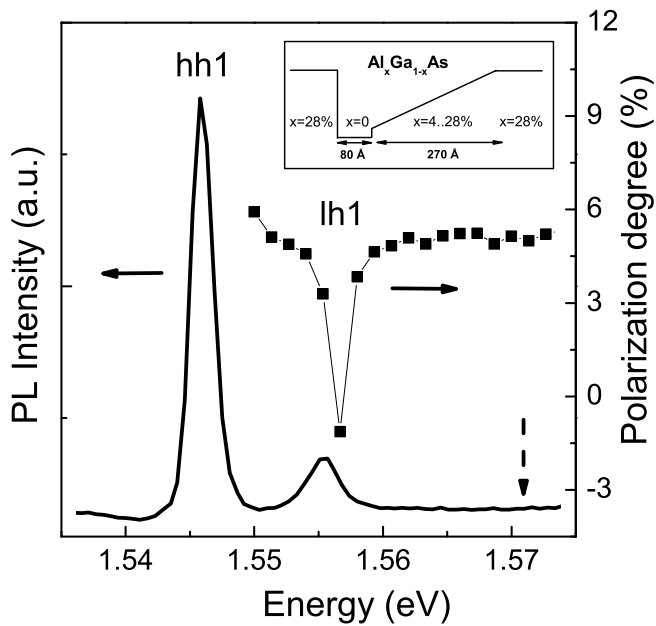}
\caption{PL spectrum under quasi-resonant excitation of the AlGaAs
QW and polarized PL excitation spectrum registered at the hh1
maximum. The dashed arrow indicates energy of excitation in the
presence of magnetic field. Inset shows a sketch of the studied
structure.}
 \label{PL}
\end{figure}

In order to observe the expected effect an asymmetrical QW was
prepared. The sample was grown by molecular beam epitaxy method on
a semi-insulating GaAs substrate along the [100] direction and
contained a 200~nm Al$_{0.28}$Ga$_{0.72}$As barrier layer, a 80~\AA
\mbox{} GaAs QW, the other sloping barrier grown with content of
Al changing from 4 to 28~\% on the length of 270~\AA, and the
barrier layer of width 200~nm. A sketch of the studied
structure is shown in the inset of Fig.~\ref{PL}. In order to
avoid oxidation of the structure there was grown a 3~nm GaAs cap
layer. Al concentration has been varied by change of the source
temperature. The growth temperature was 600$^\circ$~C. The sample
was nominally undoped.

For PL excitation we used a Ti:Sa laser pumped by???? an Ar-ion
laser. This allowed us to realize quasi-resonant excitation of QW
states. The sample was placed into a glass cryostat and was
immersed into liquid nitrogen. The cryostat was placed into an electromagnet creating a dc  magnetic field up to 0.75~T directed perpendicular to the excitation axis (Voigt
geometry). In Fig.~\ref{PL} PL spectrum of the asymmetrical QW
GaAs/AlGaAs at temperature $T=77$~K is shown. The spectrum has two
lines caused by recombination of electrons with ground states of
heavy holes (hh1) and  light holes (lh1). The same figure presents
an excitation spectrum of optical orientation signal under
registration in the main maximum of PL (hh1). Far from PL spectrum
resonances, the signal weakly depends on the excitation energy
being at the level of 5.5~\%. At the resonance, the optical
orientation signal drops sharply and became negative. This
implies that the resonance is caused by recombination of electrons
with the lh1 states.

\begin{figure}[t]
\includegraphics[width=\linewidth]{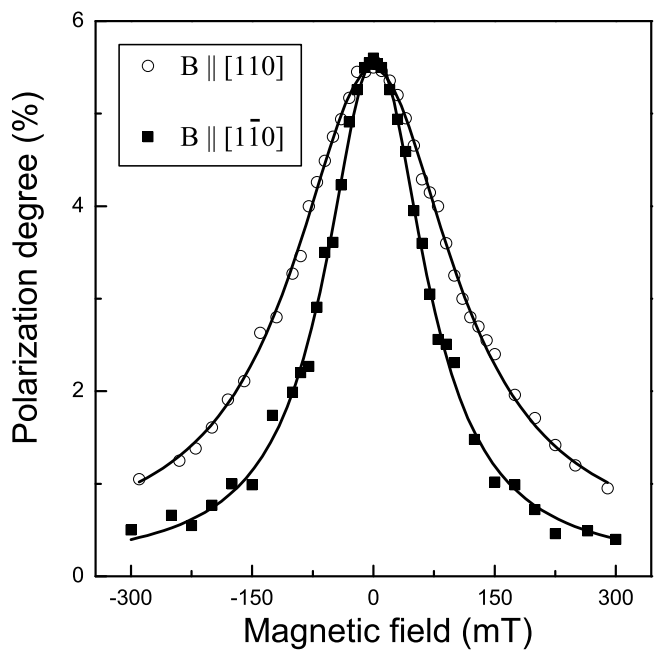}
\caption{Hanle effect measurements for two orientations of a
magnetic field in the QW plane. Solid lines represent fitting by
the Lorentz function with the halfwidths $B_+ = 0.12$~T and $B_- =
0.075$~T.}
 \label{Hanle}
\end{figure}

In Fig.~\ref{Hanle} experimental Hanle curves are presented for
two orientations of a magnetic field. Open circles and closed
squares correspond to magnetic field  directed along the axes
[110] and $[1 \bar{1} 0]$. One can see that optical
orientation is almost totally suppressed in the field 0.3~T,
however widths of two curves differ significantly. Solid lines in
Fig.~\ref{Hanle} represent fitting of the Hanle curves by the
Lorentz function. Obtained fitting parameters $B_+$ and $B_-$ for
two orientations of the field are 0.12~T and 0.075~T. This means
that Hanle linewidth anisotropy is around 60~\%.

Since $P_{circ}(0) \approx 6$~\%, the spin relaxation times
$\tau_i$ are much shorter than the radiative recombination time
$\tau_0$. This allows us to use Eq.~(\ref{B_pm}) in the analysis,
where $B_\pm$ are determined solely by the spin relaxation times.
Therefore the Hanle contour halfwidths for two field orientations
yield the spin relaxation times and the ratio $\alpha/\beta$, i.e.
relative strengths of the Rashba and Dresselhaus splittings in the
studied structure - cf.~Eq.(\ref{ratio}).

In the analysis we neglect anisotropy of the $g$-factor in the QW
plane because even if observed it does not exceed
10~\%~\cite{g_aniz}. Despite of this anisotropy also originates
from $\bm k$-linear terms in the Hamiltonian~(\ref{HSO}), its
magnitude is small due to smallness of the spin
splitting~\cite{KK}. Therefore it can be important only if the
isotropic part of the $g$-factor is close to zero. In the AlGaAs
heterostructures under study, $|g| \approx 0.35$~\cite{ivch_book},
i.e. sufficiently differs from zero, therefore its anisotropy is
inessential.

Calculation by Eq.~(\ref{B_pm}) with $|g| = 0.35$ yields
$$\tau_- = 0.8~\mbox{ns}, \:\:\:\: \tau_+ =0.3~\mbox{ns},
\:\:\:\: \tau_z =0.2~\mbox{ns},$$
and for the ratio of the spin splittings we get from
Eq.~(\ref{ratio})
$$
\left| {\alpha \over \beta} \right| \approx 4\:.
$$

Note that both [110] and $[1 \bar{1} 0]$ directions belong to the
same family of crystallographic axes. Therefore, strongly
speaking, one can determine either $|\alpha/\beta|$ or the
reciprocal value $|\beta/\alpha|$ from Hanle effect measurements.
However the Rashba splitting is usually larger than the
Dresselhaus splitting in [001] GaAs QWs. Therefore we believe that
the value given above corresponds namely to $|\alpha/\beta|$, i.e.
the Rashba splitting is about four times larger than the
Dresselhaus splitting in the studied structure. This value of
$|\alpha/\beta|$ agrees well with data on III-V
QWs~\cite{Jusserand95p4707,Miller03p076807,BIASIA}.

To summarize, electron spin relaxation anisotropy is observed in
the [001] grown QW. The anisotropy is measured by dependence of
the Hanle linewidth on magnetic field orientation in the QW plane.
It is demonstrated that the Rashba effect dominates the
Dresselhaus effect in the studied structure. Spin relaxation times
of electrons in the [001] QWs at liquid nitrogen temperature are
determined for all three spin orientations.

Financial support from RFBR and INTAS is gratefully acknowledged.
Work of L.E.G. is also sponsored by ``Dynasty'' Foundation ---
ICFPM and by Russian President grant for young scientists.

\end{document}